\documentclass[12pt, letterpaper]{article}
\usepackage[top=1in, bottom=1.5in, left=1in, right=1in]{geometry}
\usepackage[utf8]{inputenc}
\usepackage{booktabs}
\usepackage{hyperref}

\usepackage[numbers,sort&compress]{natbib}
\usepackage{amsmath,amssymb}
\usepackage{xcolor}
\usepackage{float}
\usepackage{graphicx}
\usepackage{enumitem}

\newcommand{\wfirst}{{\em WFIRST }}
\newcommand{\wfirstns}{{\em WFIRST}}
\newcommand{\lsst}{{\em LSST }}
\newcommand{\lsstns}{{\em LSST}}

\title{The Diverse Science Return from a Wide-Area Survey \\of the Galactic Plane}
\author{R.A.~Street, M.B.~Lund, S. Khakpash, M.~Donachie, W.A.~Dawson, \\N.~Golovich, L.~Wyrzykowski, P.~Szkody, T.~Naylor, M.~Penny, \\N.~Rattenbury, M.~Dall'Ora, W.I.~Clarkson, D.~Bennett, J.~Pepper, \\M.~Rabus, M.P.G.~Hundertmark, Y.~Tsapras, R.~Di Stefano, \\S.~Ridgway, M.~Liu, E.~Lin,\\ with the support of the \\
LSST Transient and Variable Stars Collaboration. }
\date{Nov 2018}

\begin{document}

\maketitle

\begin{abstract}
The overwhelming majority of objects visible to \lsst lie within the Galactic Plane.  Though many previous surveys have avoided this region for fear of stellar crowding, \lsstns's spatial resolution combined with its state-of-the-art Difference Image Analysis mean that it can conduct a high cadence survey of most of the Galaxy for the first time.  
Here we outline the many areas of science that would greatly benefit from an LSST survey that included the Galactic Plane, Magellanic Clouds and Bulge at a cadence of 2--3\,d. Particular highlights include measuring the mass spectrum of black holes, and mapping the population of exoplanets in the Galaxy in relation to variations in star forming environments.  But the same survey data will provide a goldmine for a wide range of science, and we explore possible survey strategies which maximize the scientific return for a number of fields including young stellar objects, cataclysmic variables and Neptune Trojans.
\end{abstract}

\section{White Paper Information}

\noindent{\bf Science Category:} Exploring the transient optical sky, constraining dark matter, taking an inventory of the Solar System, mapping the Milky Way\\
\noindent{\bf Survey Type Category: } Wide-Fast-Deep survey\\
\noindent{\bf Observing Strategy Category: } Integrated program with science that hinges on the combination of pointing and detailed observing strategy\\

\noindent{\bf Author's contact information:}\\

R.A.~Street, Las Cumbres Observatory, rstreet@lco.global

M.B.~Lund, Vanderbilt University, michael.b.lund@vanderbilt.edu

S. Khakpash, Lehigh University, somayeh.khakpash@gmail.com 

M.~Donachie, University of Auckland, m.donachie@auckland.ac.nz

W.A.~Dawson, Lawrence Livermore National Laboratory, will@dawsonresearch.com

N.~Golovich, Lawrence Livermore National Laboratory, golovich1@llnl.gov,

L.~Wyrzykowski, Astronomical Observatory University of Warsaw, lw@astrouw.edu.pl

P.~Szkody, University of Washington, szkody@astro.washington.edu 

T.~Naylor, University of Exeter, UK, timn@astro.ex.ac.uk

M.~Penny, Ohio State University, penny@astronomy.ohio-state.edu

N.~Rattenbury, University of Auckland, n.rattenbury@auckland.ac.nz

M.~Dall'Ora, National Institute for Astrophysics, Italy, massimo.dallora@inaf.it

W.I.~Clarkson, University of Michigan, wiclarks@umich.edu

D.~Bennett, NASA Goddard Space Flight Center, david.p.bennett@nasa.gov

J.~Pepper, Lehigh University, joshua.pepper@lehigh.edu

M.~Rabus, Pontificia Universidad Cat\'olica de Chile, markus.rabus@gmail.com

M.P.G.~Hundertmark, Universit\"{a}t Heidelberg,

markus.hundertmark@uni-heidelberg.de,

Y.~Tsapras, Universit\"{a}t Heidelberg, ytsapras@ari.uni-heidelberg.de

R.~Di Stefano, CfA, Harvard, rdistefano@cfa.harvard.edu

S.~Ridgway, NOAO, ridgway@noao.edu

M.~Liu, Institute for Astronomy, University of Hawai'i, mliu@ifa.hawaii.edu

Hsing Wen (Edward).~Lin, University of Michigan, hsingwel@umich.edu

\vspace{0.5cm}
\noindent Street, Penny and Bennett are representatives of the WFIRST Microlensing Science Investigation Team (Bennett is deputy team lead).  
\clearpage

\section{Scientific Motivation}
\label{sec:science_case}
\lsst can make outstanding contributions to an array of stellar astrophysics by 
conducting a wide-area survey of the Galactic Plane.  Here we concentrate on a selection of cases, aware of a number of complementary proposals (led by Strader et al., Dell'Ora et al., Knut \& Skcody and particularly, Lund et al.) that focus on several topics which will also benefit greatly from \lsst surveying the Galactic Plane.  

\begin{figure}[H]
\begin{centering}
\begin{tabular}{c}
\includegraphics[width=0.6\textwidth]{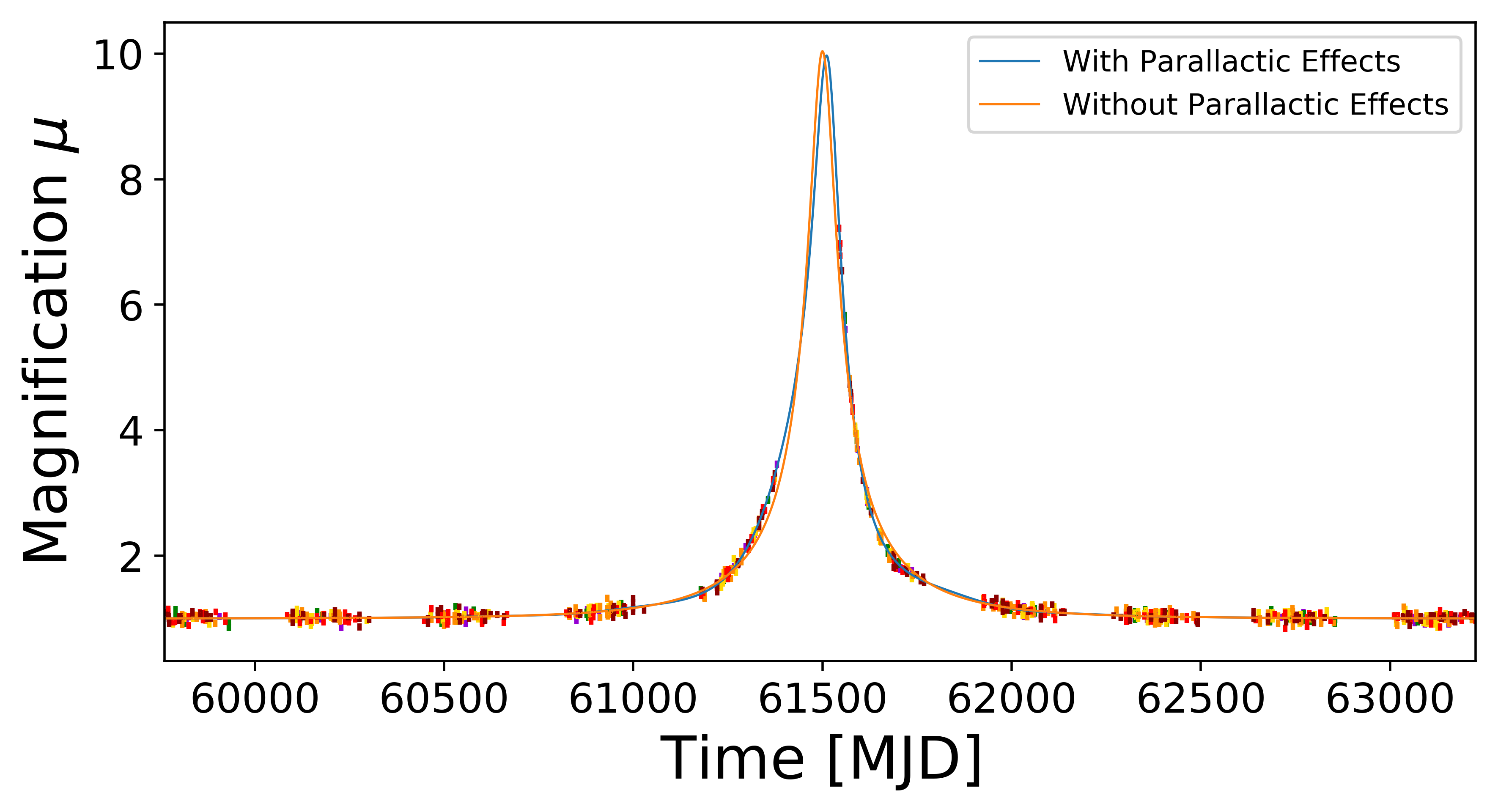}\\
\includegraphics[width=1.0\textwidth]{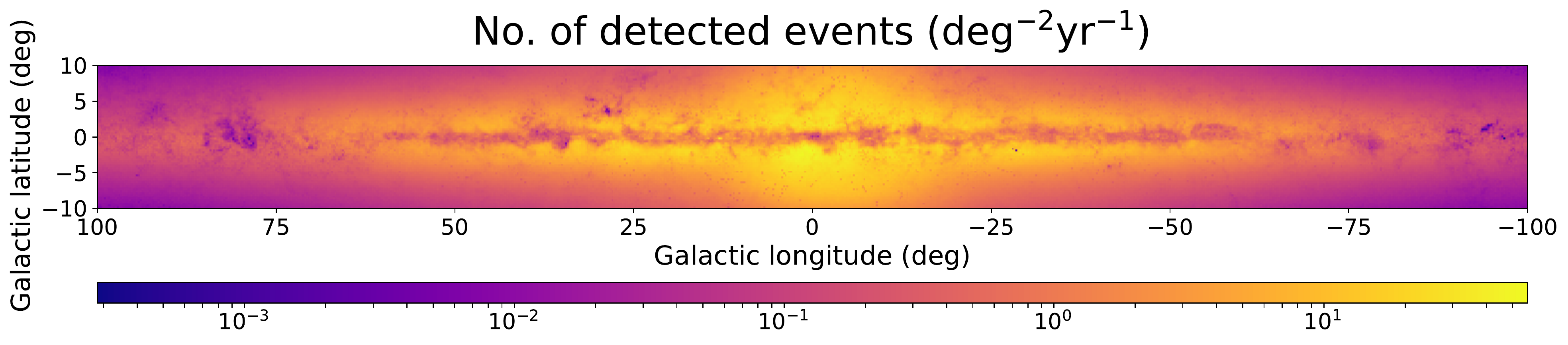}\\
\end{tabular}
\caption{(Top) Simulated lightcurve of a 40\,M$_{\odot}$ black hole lensing event observed with the cadence of the minion16A OpSim.\label{fig:bhlc}  (Bottom) Number of microlensing events detected by LSST per year per sq. deg, {\it assuming only baseline WFD coverage of the Plane.}\label{fig:map}}
\end{centering}
\end{figure}

\noindent{\bf Black Holes (BH): } 
Stellar evolution models imply that there should be millions of black holes residing in our galaxy \citep{Gould2000}. Those with masses below $\sim$20\,M$_{\odot}$ are the expected products of stellar evolution\citep{Ozel2010}, but recent gravitational wave detections suggest an unforeseen population of more massive ($>$20\,M$_{\odot}$) BHs, which result from BH-BH mergers \citep{abbott2016, Abbott2017, Belczynski2012}. These may be produced by stellar evolution, or formed in the very early Universe from the clumps of non-barionic dark matter: primordial BH (e.g., \cite{chapline1975}, \cite{bird2016}). The mass function of the BH is still very poorly known. A recent analysis of 8 years of OGLE-III photometric time-series data \citep{wyrzykowski2016} yielded 13 candidates for dark objects found via the microlensing method, indicating a continuation of mass spectrum of black holes from 3\,M$_{\odot}$ upwards.  Microlensing occurs when a foreground object (lens) passes directly between the observer and a luminous background source.  The gravity of the lens deflects light from the source with a characteristic radius, R$_{E}$, causing the source to brighten and fade as they move into and out of alignment, with a timescale, $t_{E}$ that is proportional to the lens mass (years for BHs, Fig~\ref{fig:bhlc}).  Since detection does not depend on light from the lens, the technique is a powerful means of searching for objects which are otherwise too faint to see, and the method is complementary to LIGO since it is capable of directly measuring the properties of single as well as binary, lenses. \lsst will reach fainter magnitudes than OGLE ($\sim$23 mag vs $\sim$20 mag), so we expect that by monitoring a few billion stars in the Galactic Plane it will detect hundreds of BH events.  By conducting a long duration, self-consistent survey of the Bulge, Galactic Disk and both Magellanic Clouds (LMC, SMC) \lsst will compare the populations in different local environments.  

\noindent{\bf Planetary Microlensing: } Despite outstanding discoveries from {\it Kepler} and other surveys, the vast majority of known exoplanets lie within 1\,kpc of the Sun \citep{exoplanetarchive}.  Variations in star formation, metallacity and stellar density and ages across the Galaxy mean we cannot assume that planets are ubiquitous, and comparing their occurrence rates in different populations will offer insights into their formation processes.  Microlensing can probe to much greater distances ($\leqslant$8.5\,kpc) and is sensitive to objects of all masses \citep{2001MNRAS.327..868B, 2003AcA....53..291U, 2004ApJ...606L.155B} in orbits between $\sim$1--10\,AU.  The microlensing rate for surveys outside the Bulge has been estimated \cite{SajadianPoleski2018} based on the minion\_1016 OpSim {\it with minimal coverage of Plane fields}.  They found an average rate of 15 events deg$^{-2}$ year$^{-1}$ in the disk and 400 events deg$^{-2}$ year$^{-1}$ in the Bulge.  {\bf This detection rate can be doubled if the cadence is increased from 6\,d to 2\,d (Fig~\ref{fig:map}).}  Our proposed strategy ensures regular coverage of the Magellanic Clouds.  The ``edge-on'' orientation of the SMC results in a higher probability of self-lensing (where both source and lens lie within that galaxy), raising the possibility of {\bf detecting stellar and perhaps even planetary companions in a galaxy other than the Milky Way}\cite{DiStefano2000}.  \lsst is predicted to detect 20--30 events year$^{-1}$ \cite{MrozPoleski2018}, provided the galaxy is monitored at least once every few days.  

\noindent{\bf Completing the \wfirst Bulge Survey: } \lsst will be contemporaneous with NASA's \wfirst survey of the Galactic Bulge.  The \wfirst Mission aims to discover $>$1400 microlensing planets \citep{Penny2018} by monitoring the Bulge at very high cadence.  However its survey periods last just $\sim$72\,days, with 6--12\,month inter-season gaps.  This will not only lead to anomalous features (and hence lens companions) being missed, but the incomplete lightcurves will make it difficult to reliably measure the microlensing parallax, and hence the mass and distance, of the lenses.  \lsst can significantly improve the scientific return from \wfirst simply by observing this region every few days whenever it is visible.  

\noindent{\bf Mesolensing: } R$_{E}$ is inversely proportional to the lens distance, $\sim$\,${D_L}^{-3/2}$, so nearby objects traveling at relatively high velocities are more likely to lens background stars than a similar, more distant objects \citep{2008ApJ...684...46D, 2008ApJ...684...59D}. \lsst will investigate the mass distribution of faint objects in the local neighbourhood such as low mass dwarfs, stellar remnants, and free-floating planets \citep{2008ApJ...684...59D}.  

\noindent{\bf Predicted Lensing: } \lsst proper motions will be valuable for predicting future microlensing events, as has been done from Gaia and Pan-STARRS 1 data (to shallower limiting magnitudes than \lsst), \cite{NeilsenBramich2018}.

\noindent{\bf Young Stellar Objects: } The short ($<$2 weeks) timescale variations of YSOs are thought to be dominated by magnetic-field structures locked in the rotation frame of the star and for scheduling reasons most previous YSO studies have probed only the innermost few stellar radii of the star and disk.  However, planets are thought to form at radii of a few AU in protoplanetary disks \citep{WilliamsCieza2011}, raising vertical structures and perhaps warps in the disk. Occultations by these structures can yield information about the physics of the disc, and hence planet formation, but data with sufficient baseline to detect them are sparse.  Disk occultation systems are rare, so only a very wide-area survey will produce a significant sample.  \lsstns's deep limiting magnitude will probe variability in YSO populations where is has not been explored, including below the deuterium burning limit in nearby star-forming regions. 

\noindent{\bf Cataclysmic Variables (CVs): } CVs, (close binaries with a white dwarf accreting from a late main-sequence star), are the most common end-product of binary star evolution, and their correct numbers are necessary for a complete understanding of stellar evolution. They also constitute possible progenitors for SN Ia.  Dwarf novae are most easily discovered when they undergo outbursts caused by disk instability-caused with rise times of 0.5-1 day, lasting 2-30 days), but these occur unpredictably, at intervals of 1 week to 30 years.  \lsstns's long baseline of observations over a wide region of the galaxy will discover dwarf novae via outbursts caused by disk instabilities. 

\begin{figure}[H]
\begin{centering}
\begin{tabular}{cc}
\includegraphics[width=0.35\textwidth]{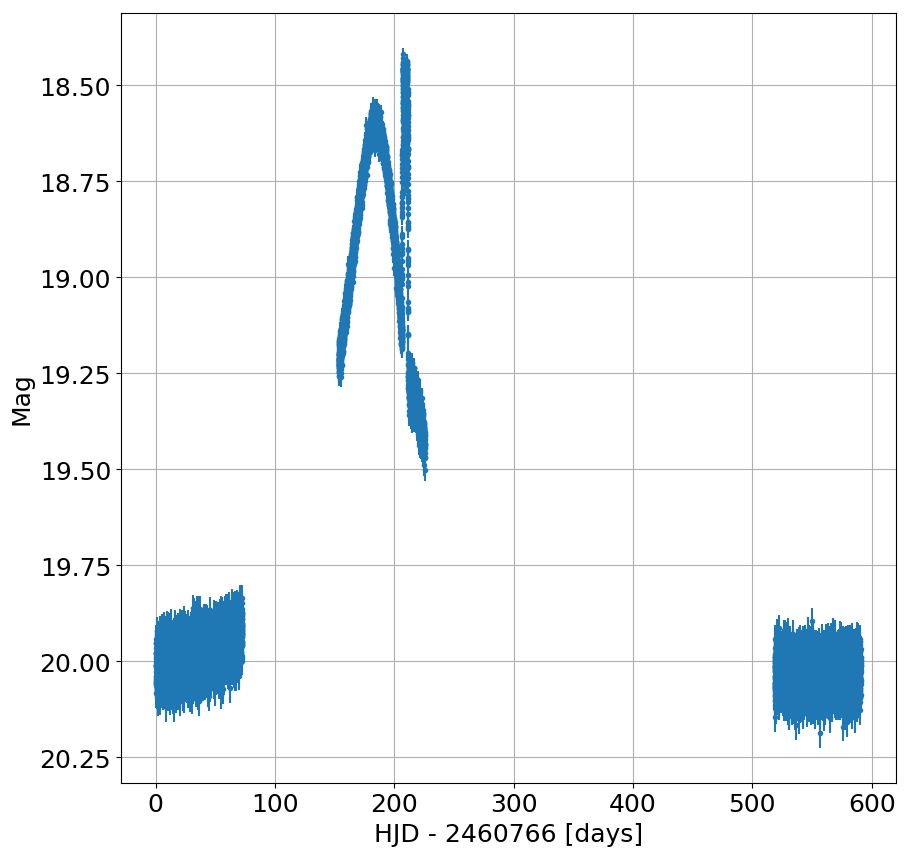}&
\includegraphics[width=0.60\textwidth]{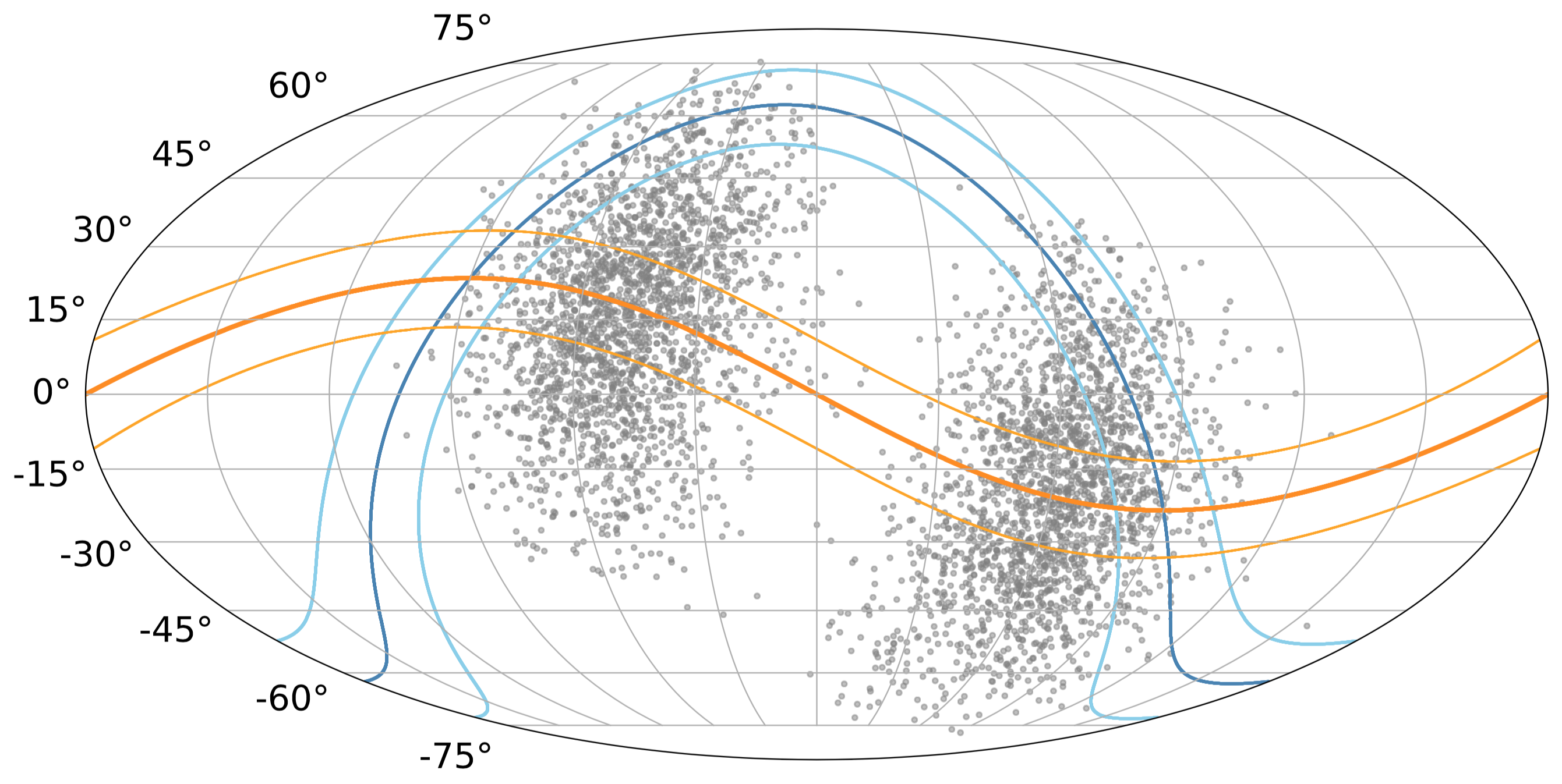}
\end{tabular}
\caption{(Right) Simulated lightcurve of a planetary microlensing event observed by \wfirstns, showing the $\sim$72\,d survey `seasons' and the extended gaps between them.  \lsst will provide important coverage during these data gaps. \label{fig:wfirstlc} (Right)
Expected sky positions of the predicted population of Neptune trojans in 2022. \label{fig:trojans}}
\end{centering}
\end{figure}

\noindent{\bf Neptune Trojans: } Trojans are asteroids that orbit at the stable Lagrangian points, 60$^{\circ}$ ahead and behind the major planets; currently, 22 are known orbiting with Neptune \cite{MPC}.  Uncovering their true population would help us to understand their origins and dynamics, but they are faint ($r$\,$\sim$22.5-24.5) with relatively low proper motions ($\sim$2$^{\circ}$ year$^{-1}$).  During \lsstns's main science operations, one of the trojan groups will lie close to the Galactic Plane (Fig.~\ref{fig:trojans}), of whom $\sim$10 are expected to be brighter than $r$=22.5mag.  

\vspace{.6in}

\vspace{.4in}

\vspace{.6in}

\section{Technical Description}

\subsection{High-level description}
\begin{footnotesize}
{\it Describe or illustrate your ideal sequence of observations.}
\end{footnotesize}

We propose that LSST conduct a survey of the Galactic Plane, Bulge and Magellanic Clouds at a 2--3\,d cadence, but using a reduced filter set to minimize the impact on survey cadence over the rest of the sky.  Since the yield of galactic science can be optimized (to first order) by maximizing the number of stars covered, we propose that the survey footprint (and cadence, if necessary) be a function of the spatial density of stars within a region -10.0 $\leqslant b \leqslant $+10.0$^{\circ}$.  

\lsstns's rapid data reduction and alert system will enable it to act as a discovery engine for microlensing events and CVs entering outburst, but to effectively alert on these transient phenomena, each field must be sampled with a regular cadence of $\sim$1-2 visits per day.  Alerts will be identified and selected from brokers such as ANTARES \citep{ANTARES} in real-time, and prioritized by automated algorithm (e.g. \citep{Hundertmark2018}).  Follow-up observations will be executed across a network of ground-based telescopes of different apertures, depending on the brightness of the target.  This alert and follow-up strategy has been well established in the course of previous surveys (e.g. \citep{Yee2014, Street2016}).  This will enable us to characterize the shorter timescale ($\sim$hours--days) anomalous features of the lightcurves which betray the companion objects to the lens.  

A single filter normally dominates in microlensing observations, since the phenomenon is wavelength-invarient to first order.  However, time series observations in at least two  (ideally 3) filters are still required to deblend the color of the source star, which is needed in order to estimate its spectral type and hence its angular diameter.  This is a necessary component in the modeling process, required in order to determine the physical properties of the lens.   Multi-filter lightcurves can be combined to analyze the event, and moreover provides a wealth of additional science.   We propose to restrict the filters used to $g,r,i,z$, to minimize overheads.  Multi-filter data are also important in the detection of CV outbursts since the targets are relatively blue and to characterize long-term YSO variability.  

We propose to image all pointings using a ``paired-$i$'' strategy, implemented at intervals of 2-3\,d.  On a night when a given field is to be surveyed, the first visit to the field would be in $i$- band.  If a second visit is possible that night within $>$1\,hr and $<$4\,hrs, then it would be taken in one of $g,r,z$ filters, where the filter selection alternates through the set each day.  Each visit would consist of a single, 30\,s exposure. This cadence should be maintained for as long as each field is visible.  

\vspace{.3in}

\subsection{Footprint -- pointings, regions and/or constraints}
\begin{footnotesize}{\it Describe the specific pointings or general region (RA/Dec, Galactic longitude/latitude or 
Ecliptic longitude/latitude) for the observations. Please describe any additional requirements, especially if there
are no specific constraints on the pointings (e.g. stellar density, galactic dust extinction).}
\end{footnotesize}

The footprint should cover as much of the Galactic Plane as possible between -10.0 $\leqslant b \leqslant $+10.0$^{\circ}$ and -85.0 $\leqslant l \leqslant$+85.0$^{\circ}$, with priority and cadence given in proportion to the number of stars in each field, with additional pointings covering the Magellanic Clouds.  
This automatically gives weight to the Bulge and other regions of high star count, while low priority is given to the regions of highest extinction.  

\subsection{Image quality}
\begin{footnotesize}{\it Constraints on the image quality (seeing).}\end{footnotesize}

The proposed survey region includes fields of high crowding, where stellar Point Spread Functions (PSFs) overlap.  While some degree of overlap is inevitable and can be taken into account with existing analysis techniques, photometric measurements become significantly less precise during periods of poor seeing.  We therefore request observations in conditions of $<$2\,arcsec.  

\subsection{Individual image depth and/or sky brightness}
\begin{footnotesize}{\it Constraints on the sky brightness in each image and/or individual image depth for point sources.
Please differentiate between motivation for a desired sky brightness or individual image depth (as 
calculated for point sources). Please provide sky brightness or image depth constraints per filter.}
\end{footnotesize}

This survey has minimal constraints on sky brightness, since it prioritizes regularity of cadence over limiting magnitude, and co-added images are not essential for the science.   The only exception is to avoid occasions when the Moon passes through the galactic plane.  Since LSST's scheduler already includes a lunar-avoidance metric, this will be sufficient for our purposes.

The limiting magnitude expected per bandpass for a single 30\,s exposure is estimated below, taking the extinction into account. 

\begin{table}[h!]
\centering
\begin{tabular}{lllll}
\hline
             & Sky brightness & Limiting magnitude \\
\hline
SDSS-$g$     &  All           &  23.0\\
SDSS-$r$     &  All           &  23.3\\
SDSS-$i$     &  All           &  22.9\\
SDSS-$z$     &  All           &  22.3\\
\hline
\end{tabular}
\end{table}

\subsection{Co-added image depth and/or total number of visits}
\begin{footnotesize}{\it  Constraints on the total co-added depth and/or total number of visits.
Please differentiate between motivations for a given co-added depth and total number of visits. 
Please provide desired co-added depth and/or total number of visits per filter, if relevant.}
\end{footnotesize}

Most of our science cases do not require the co-addition of images, with the possible exception of Neptune Trojans.   Ten objects are expected to be brighter than $r$=22.5mag.

The total number of visits to all pointings in this survey depends on the survey cadence implemented, which may vary depending on the field (please see section below on estimated time requirements).  The table below provides estimated total visits per year for the entire survey region, assuming different uniform cadences.

\begin{table}[h!]
\centering
\begin{tabular}{lcc}
\hline
                     & $|b|\pm10^{\circ}$ & $|b|\pm5^{\circ}$ \\
Strategy             & N visits per year       &  N visits per year \\
\hline
Uniform 2\,d cadence &   23,766             & 17,748 \\
Uniform 3\,d cadence &   15,844             & 11,832 \\
\hline
\end{tabular}
\end{table}

\subsection{Number of visits within a night}
\begin{footnotesize}{\it Constraints on the number of exposures (or visits) in a night, especially if considering sequences of visits.  }
\end{footnotesize}

When the visibility of a field is between 0.5 -- 4\,hrs per night, a single visit should be conducted. When the visibility of the field is $>$4\,hrs per night, two visits should be made, the first in $i$-band and the second in one of $g,r,z$.  The requirement for paired observations is driven by the need to obtain color lightcurves for YSOs, although it is highly beneficial for microlensing and CV science also, because it means a higher probability of catching those events on the rise (or during caustic crossings) at earlier times.  

\subsection{Distribution of visits over time}
\begin{footnotesize}{\it Constraints on the timing of visits --- within a night, between nights, between seasons or
between years (which could be relevant for rolling cadence choices in the WideFastDeep. 
Please describe optimum visit timing as well as acceptable limits on visit timing, and options in
case of missed visits (due to weather, etc.). If this timing should include particular sequences
of filters, please describe.}
\end{footnotesize}

Regular time sampling of target lightcurves is essential for all our science cases, but it is also critical that it be conducted over a long time baseline.  This is particularly important for lensing by BH candidates, for which typical timescales range between 100--300\,d. Regular sampling of microlensing lightcurves is vital in order to accurately measure the annual microlensing parallax signature, which manifests as a skew in the lightcurve over the course of the event.  The more frequent the observations, the better constraints on the mass of the lens.  However, our simulations indicate that a sampling of $<$40\,d will be enough to yield a BH lens mass with 10-20 \% accuracy.

Regular cadence is also required in order to accurately alert on new lensing events in time for follow-up to be conducted (i.e. on the early part of the rising lightcurve); the same is true for CVs in outburst.  Stellar-mass lens event have timescale $\sim$5-100\,d, with a distribution that peaks around 20\,d.  At least 3 datapoints are needed to accurately distinguish between CVs, SNes, flare and microlensing events based on the shape of their rising lightcurve with a reasonable rate of false positives.  Microlensing is most sensitive to planets near the peak of the event, with a median $t_{E}\sim20$--$30$\,d. Therefore 3 days is the absolute lowest cadence that you can guarantee detection before the peak for tE$\geqslant$10\,d, in time for follow-up observations to be conducted.  The minimum cadence required to detect long-term trends in YSOs is $\sim$once per week.  Dwarf novae have typical rise times of 0.5--1 day, with outbursts lasting 2--30 days, that occur unpredictably, at intervals of 1 week to 30 years.

Gaps in the time sampling negatively impact the study of the most scientifically interesting microlensing events -- binary lenses -- which exhibit short ($\sim$few days) `spikes' in magnification as well as trough and curving features.  Delivering cadence sufficient to adequately constrain the timing and change in brightness during these features is essential to the characterization of the lens.  A rolling cadence would provide better coverage and hence better characterization of binary lightcurves when in a `favored' year, at the expense of significantly poorer coverage in other years.  Since microlensing events are transients and anomalies can occur at any time, this reduces the sample of anomalous events detected overall.  

A rolling cadence would negatively impact our abilty to detect long-term trends in YSO stars, as well as the probability of detecting outbursting CVs.  That said, a rolling cadence would make it easier to constrain the orbital arcs of Neptune trojans by providing higher cadence lightcurves, provided the Plane fields were prioritized at the start of \lsstns's science operations.  

Since microlensing and CV outbursts are transient events, our total yield is maximized by maintaining a regular cadence throughout \lsstns's operational lifetime.  Observations early in \lsstns's operations will be vital in order to produce a deep catalog of all variable stars within the field.  This will subsequently be used to filter out possible sources of false positive alerts.  
 
In the event of data gaps due to bad weather, observations should restart as soon as possible in the $i$ filter. The Moon passes through the Plane multiple times per year, during which \lsst should concentrate on observing those fields outside its lunar exclusion zone.  A further optimization may be made by imaging only in $r, i, z$ filters during these periods.  

\subsection{Filter choice}
\begin{footnotesize}
{\it Please describe any filter constraints not included above.}
\end{footnotesize}

Our preferred filter is $i$, which offers the optimum combination of telescope throughput, detector sensitivity and target star intrinsic brightness while providing as good a match as possible with other datasets taken in the Galactic Plane.  We propose to use $g,r$ data for detecting CVs in outburst and trends in YSOs, and $g,r,i,z$ for measuring star colors and time series observations in regions of higher extinction.  

We do not request $u,y$ filters, because the high extinction of our fields and the low detector QE are not ideal and there is insufficient scientific benefit to justify the additional overheads.  

\subsection{Exposure constraints}
\begin{footnotesize}
{\it Describe any constraints on the minimum or maximum exposure time per visit required (or alternatively, saturation limits).
Please comment on any constraints on the number of exposures in a visit.}
\end{footnotesize}

Our requested exposure times are determined by the need to reach limiting magnitudes $>$22\,mag in all passbands.  The majority of microlensing source stars have $i > $16\,mag (based on existing data from ground-based surveys), and so fit well within \lsstns's dynamic range for a 30\,s exposure.  The detection of Neptune trojans requires us to reach $r\sim $23\,mag.  
Follow-up observations will be obtained from other ground-based resources for priority targets that are magnified to brightnesses which would saturate in \lsstns's images.  

\subsection{Other constraints}
\begin{footnotesize}
{\it Any other constraints.}
\end{footnotesize}

We note strong synergies between this proposal and the Deep Drilling Field proposed to cover the WFIRST Microlensing Survey footprint in the Galactic Bulge, since this pointing will automatically be included in this survey thanks to the region's high stellar density.  Ideally, this wide-area survey would begin immediately after LSST commissioning is complete and continue for the 10\,yr duration of the WFD, to maximize the yield of long timescale, BH events, and to ensure the best possible phase coverage of all forms of stellar variability.  At minimum, we request that it begin at least 1\,yr before the start of the WFIRST survey ($\sim$2026), and continue for at least the same years as WFIRST surveys the Bulge.  

A natural product of LSST's Galactic Plane survey will be a comprehensive catalog of all variable stars within the survey region, which (in addition to their intrinsic scientific value) would be an important input catalog for the WFIRST Mission, particularly if (as currently under planning) it provides daily alerts of ongoing microlensing events.  The catalog would enable the large majority of false positives to be entirely avoided, thereby considerably enhancing the productivity of the mission as well as follow-up observations of LSST alerts.  

\subsection{Estimated time requirement}
\begin{footnotesize}
{\it Approximate total time requested for these observations, using the guidelines available at \url{https://github.com/lsst-pst/survey_strategy_wp}.}
\end{footnotesize}

There are various possible permutations of this wide-area survey which can be made, such as restricting the footprint (e.g. to $|b|\pm5^{\circ}$ rather than $|b|\pm10^{\circ}$), or applying different cadence (2 vs. 3\,d) to different sub-regions (e.g. higher cadence for the Magellanic Clouds, while a lower cadence for $|b|>5^{\circ}$, for example) in proportion to the number of stars in those fields.  We are conducting further investigations to determine the optimal trade-off, so the following is given as a rough guide only. \\

\noindent Single visit per field conducted during low-visibility windows ($<$4\,hrs per night), 1$\times$30\,s exposure including overheads = 162\,s.\\
Paired visit per field for during high-visibility windows ($>=4$\,hrs per night), 2$\times$30\,s exposures including overheads = 324\,s.\\
On average, each field has low-visibility for 90\,nights per year and high-visibility for 114\,nights per year.  \\
The Galactic Plane is covered by 233 \lsst pointings within the Galactic Plane region 5$\leqslant$RA$\leqslant$21\,hrs and $|b|\pm10^{\circ}$, or by 174 pointings for $|b|\pm5^{\circ}$.\\

\begin{table}[h!]
\centering
\begin{tabular}{lcc}
\hline
                     & $|b|\pm10^{\circ}$ & $|b|\pm5^{\circ}$ \\
Strategy             & hrs per year       &  hrs per year \\
\hline
Uniform 2\,d cadence &   1667             & 1245 \\
Uniform 3\,d cadence &   1111             & 830 \\
\hline
\end{tabular}
\end{table}

\vspace{.3in}

\begin{table}[ht]
    \centering
    \begin{tabular}{l|l|l|l}
        \toprule
        Properties & Importance \hspace{.3in} \\
        \midrule
        Image quality &   2  \\
        Sky brightness &  3\\
        Individual image depth &  1 \\
        Co-added image depth &  3 \\
        Number of exposures in a visit   &  2 \\
        Number of visits (in a night)  & 1  \\ 
        Total number of visits & 2  \\
        Time between visits (in a night) & 1 \\
        Time between visits (between nights)  &  1 \\
        Long-term gaps between visits & 1\\
        Other (please add other constraints as needed) & \\
        \bottomrule
    \end{tabular}
    \caption{{\bf Constraint Rankings:} Summary of the relative importance of various survey strategy constraints. Please rank the importance of each of these considerations, from 1=very important, 2=somewhat important, 3=not important. If a given constraint depends on other parameters in the table, but these other parameters are not important in themselves, please only mark the final constraint as important. For example, individual image depth depends on image quality, sky brightness, and number of exposures in a visit; if your science depends on the individual image depth but not directly on the other parameters, individual image depth would be `1' and the other parameters could be marked as `3', giving us the most flexibility when determining the composition of a visit, for example.}
        \label{tab:obs_constraints}
\end{table}

\subsection{Technical trades}
\begin{footnotesize}
{\it To aid in attempts to combine this proposed survey modification with others, please address the following questions:}
\begin{enumerate}
    \item {\it What is the effect of a trade-off between your requested survey footprint (area) and requested co-added depth or number of visits?}
    
    Overall, reducing the survey area will impact the number of the different categories of targets that we discover.  However, there are optimisations which can be made to minimize the time required for the survey while maintaining the majority of the science.  In particular, we recommend prioritizing those fields with the highest number of stars.  
    
    \item {\it If not requesting a specific timing of visits, what is the effect of a trade-off between the uniformity of observations and the frequency of observations in time? e.g. a `rolling cadence' increases the frequency of visits during a short time period at the cost of fewer visits the rest of the time, making the overall sampling less uniform.}
    
    Much of our science focuses on the long-timescale trends (microlensing, YSOs) in our targets, so non-uniform sampling is likely to negatively impact our program overall.  Since many of our phenomena of interest are transients (microlensing, CVs in outburst), new events occur all the time, so a reduced cadence at any time reduces the overall yield relative to a uniform cadence.  
    
    The exceptions to this are that rolling cadence would achieve better characterization of microlensing anomalies in years of favorable coverage (at the expense of increasing the rate of missed detections in years or low coverage).  Since this characterization is likely to be provided by follow-up facilities, this is not a compelling argument.  The other exception is that Neptune Trojans would become easier to detect, provided the rolling cadence favored those fields at the start of \lsst science operations.  
    
    \item {\it What is the effect of a trade-off on the exposure time and number of visits (e.g. increasing the individual image depth but decreasing the overall number of visits)?}
    
    Our primary concern is the regular sampling of target lightcurves.  A practical minimum sampling would be every 3\,d, but this would constrain us to focus on the longer duration microlensing events only, with a much lower probability of detecting the anomalies that lead to the characterization of stellar, brown dwarf and planetary companions.  
    
    Reaching deeper limiting magnitudes will increase the number of fainter targets, which would be beneficial for most science cases particularly the detection of more Neptune Trojans, provided the cadence remains sufficiently high to detect enough of their orbital arc.  
    
    Pushing towards deeper limiting magnitudes in the Galactic Plane is somewhat less beneficial overall since practical limits on depths are imposed by the extinction and confusion limits. 
    
    \item {\it What is the effect of a trade-off between uniformity in number of visits and co-added depth? Is there any benefit to real-time exposure time optimization to obtain nearly constant single-visit limiting depth?}
    
    Optimizing the exposure time to reach a constant limiting magnitude would improve the number of Neptune Trojans that can be confirmed, since they are close to the limiting magnitude.  In all other science cases, the benefits are rather less clear, and cadence remains our chief concern for the reasons explained above. 
    
    \item {\it Are there any other potential trade-offs to consider when attempting to balance this proposal with others which may have similar but slightly different requests?}
    
    Much if not all of our science can be done in $i$ or $r$ filters, so the choice of primary filter is somewhat flexible, though $i$ penetrates the extinction better.  As described above, much of our science can be achieved with a slightly lower cadence, but the best optimization would trade overall survey area for regular cadence of the remaining fields, provided the remaining fields are those with the highest numbers of stars. Alternatively, we could cover the most densely populated fields at the cadence requested here, while monitoring the remaining fields with lower cadence.  \\
    
    We note strong synergies between this proposal and the white paper lead by Lund et al.  Discussions between our teams have indicated that were Lund et al. is primarily concerned with the total overall numbers of observations, this proposal is concerned with the cadence.  Nevertheless, we believe that the strategies can be successfully merged.  
    
\end{enumerate}
\end{footnotesize}
\pagebreak
\section{Performance Evaluation}

The guiding criteria for our strategy are 1) maximizing number of stars monitored within the Galactic Plane, with selected fields (SMC, LMC, WFIRST footprint) given priority and 2) regularity of survey cadence in at least one ($i$) filter.  To this end, we propose the following metrics, $M$, for which hotlinks are given below (bold font) to public Github repositories.  \\

\noindent {\bf Survey Footprint Metrics}\\
The rate of microlensing is directly proportional to the number of stars $N_{stars}$ in a field, $f$, so our scientific yield can be optimized by maximizing $M = \sum_{f}( p(f)\rm{N_{stars}(f)} ),$ where priority $p(f) = 1.0$ for fields including the SMC, LMC and WFIRST Bulge Survey footprint, $p(f) = 0.75$ for other fields with $|b| < 10^{\circ}$ and $p(f) = 0.0$ elsewhere. $N_{stars}$ can be calculated using the \href{https://github.com/LSST-nonproject/sims\_maf\_contrib/tree/master/mafContrib/StarCounts/}{StarCounts} metric.  
Since the microlensing event rate is $\sim$10$^{-7}$--10$^{-6}$ star$^{-1}$ year$^{-1}$ in our proposed fields, the threshold for a minimally-useful survey requires $M>$10$^{7}$.  \\
The \href{https://github.com/LSST-TVSSC/software_tools/blob/master/SpacialOverlapMetric.py}{\bf SpacialOverlapMetric} has been written to evaluate the degree of spatial overlap between a given OpSim and specific defined regions, such as the WFIRST footprint, SMC and LMC (where we require that the overlap be 100\%).\\

\noindent {\bf Cadence Metrics}\\
It is important to ensure observations of these fields are obtained regularly, and for long periods of time.  To evaluate this, we recommend the following
\begin{itemize}[leftmargin=*]
\itemsep0em 
    \item Minimize the \href{https://github.com/lsst/sims_maf/blob/master/python/lsst/sims/maf/metrics/cadenceMetrics.py}{\bf UniformityMetric (vectorMetrics module)}: Summed over all fields within the region described above, this metric is zero for perfectly uniform observations.
    \item Maximize the  \href{https://github.com/LSST-nonproject/sims_maf_contrib/blob/master/mafContrib/campaignLengthMetric.py}{\bf CampaignLengthMetric}: Also summed over the survey region, this metric ensures that observations persist over sufficiently long timescales.  For BH lensing and YSOs in particular, the longer the better, but a minimally-useful threshold would be $>$300\,days. 
    \item Minimize the inter-night gap between 95\% of successive observations of a given field.  This should be minimized for all fields within the proposed footprint, $M = \sum_{f} t_{gap,95\%}(f)$.  The threshold for this metric depends on the science case: BH lensing requires $<$40\,days in the Galactic disk region, i.e. $|b|<5$ deg and $|l|>15$ deg, while stellar and planetary lensing requires $<$3\,days for the whole region and $<$2\,days for the high priority regions (Bulge, SMC, LMC).  
    \item Maximize the \href{https://github.com/rachel3834/sims_maf_contrib/blob/master/mafContrib/CadenceOverVisibilityWindowMetric.py}{\bf CadenceOverVisbilityWindowMetric}: Obtaining observations of the WFIRST Survey footprint immediately prior to the start of mission operations, and during its inter-season gaps, will also be important.  This metric, $$M = [\sum_{\rm nights} (N_{\rm{visits, actual}} / N_{\rm{visits, desired}})]/N_{\rm{filters}},$$ compares the actual number of visits (in a given filter set) with the maximum possible number of visits, calculated from the visibility window of the field for the given start and end dates, and desired cadence.  Its value is 1.0 if all visits are made.  For example, over a 4-night span, if a field is visible long enough to have two visits in a given night, then four visits would ideally be made (twice on 2 nights out of the 4, in 3 filters), but the minimum scientifically-useful cadence would be once every 3\,d, giving $M = [1 / 4]/3 = 0.083$.
    \item Maximize the \href{https://github.com/Somayeh91/sims_maf_contrib/blob/master/mafContrib/numObsInSurveyTimeOverlap.py}{\bf numObsInSurveyTimeOverlap}: This metric computes the number of observations within specific time periods, e.g. during WFIRST seasons.  
    \item Minimize the \href{https://github.com/Somayeh91/sims_maf_contrib/blob/master/mafContrib/IntervalsBetweenObs.py}{\bf IntervalsBetweenObs}: This metric computes the mean and median interval between observations within specific time periods, and (as described above), different thresholds apply to different science cases. 
    \item Maximize the \href{https://github.com/lsst/sims_maf/blob/master/python/lsst/sims/maf/metrics/cadenceMetrics.py}{\bf NRevisitsMetric}: summed over all fields in the survey with a $\delta T$=240\,mins.  The minimum-useful threshold are given in the table in section 3.5 under 3\,d cadence. 
\end{itemize}

Achieving truly {\it perfect} cadence has the negative consequence of strong aliases in the phase coverage of periodic variables, so we note that it is unnecessary to repeat visits to a given field precisely at the indicated cadence, but rather that the timing of the visits within a night is flexible by $\pm$4hrs, provided gaps in the sequence are kept to a minimum.  \\

\noindent {\bf Alert Metrics}\\
For those targets which will require follow-up observations for full characterization, we recommend that the \href{https://github.com/LSST-TVSSC/software_tools/blob/master/EventTriggerMetric.py}{\bf EventTriggerMetric} be maximized using parameters delmin=4\,hrs and delmax=5\,days, set by the typical timescales of the events.  This will ensure that sufficient \lsst data are obtained for reliable alerts.  It is difficult to set a single minimum threshold for this metric, since at least some of our science will be possible without follow-up to alerts.  Maximising the value however will ensure greater return from planetary and stellar lensing.  \\

\noindent{\bf Figures of Merit, $FoM$}\\
The metrics described above can be combined to define FoM that relate directly to scientific yield.  Most importantly, $FoM = A(f) C(f) \tau(f),$ where $A(f)$ represents the survey footprint area metric, $C(f)$ the cadence over visibility window metric, and $\tau(f)$ the campaign length metric.  \\

\noindent{\bf Observational Criteria}\\
The Moon will cause unavoidable interruptions to our proposed cadence, since it passes through the Galactic Plane a number of times a year.  The lunar avoidance metric built into the \lsst scheduler should be sufficient.  

\vspace{.6in}

\section{Special Data Processing}
\begin{footnotesize}
{\it Describe any data processing requirements beyond the standard LSST Data Management pipelines and how these will be achieved.}
\end{footnotesize}

We believe the data acquired by our proposed survey will conform to the requirements of the LSST Data Management pipeline, and that it will be possible to process the data in the same manner as the data from the WFD.  

\section{Acknowledgements}
This work developed partly within the TVS Science Collaboration and the author acknowledge the support of TVS in the preparation of this paper. The authors acknowledge support from the Flatiron Institute and Heising-Simons Foundation for the development of this paper.

\section{}

\end{document}